# Highly robust and efficient metal-free water cup solid-liquid triboelectric generator for mechanical energy harvesting and ethanol detection

Kequan Xia[1], Min Yu[2*]

[1]Department of Materials Science and Engineering, National University of Singapore, Singapore 117575, Singapore;

[2]Department of Mechanical Engineering, Imperial College London, SW7 2AZ, London, United Kingdom.

*Corresponding author: m.yu14@imperial.ac.uk.

*Abstract*—Recently, low-frequency mechanical energy harvesters based on solid-liquid contact electrification have garnered widespread attention for their unique advantages in wear resistance, high charge transfer efficiency, and novel insights into electron-ion interactions at the solid-liquid interface, particularly in material identification. Hence, we designed an robust and efficient water cup triboelectric nanogenerator (WC-TENG) that only uses ordinary drinking water and plastic water cups as primary materials, achieving high-efficiency power output while eliminating the need for metal electrodes and effectively addressing the issue of corrosion in generator components. Experimental results indicate that, at an operating frequency of 2 Hz, the WC-TENG generates an open-circuit voltage ($V_{oc}$) of 249.71 V, a short-circuit current ($I_{sc}$) of 4.21 µA, and a transferred charge ($Q_{sc}$) of 188.85 nC. The WC-TENG demonstrates long-term stability and reliability, maintaining stable voltage output over 1500 s. Moreover, the WC-TENG maintains stable performance under high humidity conditions, and its output enhances with increasing temperature, underscoring its robustness and adaptability for diverse environmental applications. Furthermore, the introduction of ethanol disrupts the potential balance at the solid-liquid interface by impeding electron transfer and reducing the WC-TENG's electrical output, but as the ethanol volatilizes, the device gradually returns to its original potential state, demonstrating its potential as a selective ethanol sensor. This design not only advances the development of corrosion-resistant, high-performance energy harvesters but also opens up new possibilities for low-cost, sustainable, and environmentally adaptable sensing technologies.

## 1. Introduction

The extensive reliance on non-renewable energy sources, such as fossil fuels, has led to severe environmental pollution and energy shortages, posing a significant threat to the sustainable development of human society [1]. As a result, there has been a growing emphasis on the exploration and adoption of green energy alternatives. A key phenomenon in this context is contact electrification, also known as triboelectrification, which occurs at the interface between two materials when they come into contact and subsequently separate, leading to the generation of triboelectric charges. The origin of these charges is determined by the nature of the materials and specific experimental conditions, primarily involving the transfer of ions, electrons, or material fragments [2]. In 2012, this fundamental principle was harnessed to develop a triboelectric nanogenerator (TENG), an efficient mechanical energy harvesting device based on triboelectric and electrostatic induction principles, offering a novel and sustainable approach to environmental energy harvesting [3, 4]. Notably, TENGs exhibit significant potential for deployment in extreme environments due to their ability to efficiently harvest mechanical energy from a wide range of triboelectric materials [5, 6], versatile structural designs [7], and multiple operational modes [8-11]. Their adaptability makes them well-suited for functioning as self-powered sensors in real-time monitoring applications [12]. TENGs are capable of efficiently capturing mechanical energy from challenging environments where other technologies often fall short, such as micro-vibrations on cross-sea bridges [13, 14], low-frequency ocean waves [15, 16], water droplet [17], ice [18], and even potential applications under Martian conditions [19]. Besides, TENG device also has self-powered sensing function, which can sense various mechanical movements, environmental humidity [20], temperature [21], and material identification [22]. The range of triboelectric materials used in TENG devices is diverse, including polymers [23, 24], metals [25, 26], and composites [27, 28], each chosen for their specific properties. However, conventional solid triboelectric materials often face challenges such as wear and degradation during extended use, which limits their efficiency and lifespan. To overcome these issues, the solid-liquid triboelectric nanogenerator (SL-TENG) has been developed, offering unique advantages in wear resistance and high efficiency in harvesting low-frequency mechanical energy [29-32]. The SL-TENG utilizes liquid triboelectric materials that provide a self-renewing interface, reducing friction-related wear and enabling consistent contact with solid surfaces. This design enhances durability and allows for flexible and adaptable operation in various environments, particularly in harsh or variable conditions where traditional solid materials would fail. SL-TENGs represent a significant advancement in energy harvesting technology, combining long-term durability with the efficiency needed for capturing low-frequency energy, making them ideal for sustainable energy applications and self-powered sensor networks [33, 34].

Numerous water-based SL-TENGs have been developed to date, showcasing significant potential in efficiently converting mechanical energy into electrical energy. For example, in 2018, Pan et al. constructed a U-tube TENG operating in liquid-solid mode to investigate the impact of various liquid properties on the device's output performance [35]. Wu et al. designed a versatile and high-performance water-tube-based TENG by encapsulating deionized water (DI) within a fluorinated ethylene-propylene tube, achieving an exceptionally high volumetric output charge density of 9 mC/m³ at a frequency of 0.25 Hz [36]. You et al. developed an equivalent circuit model with governing equations for a water-solid mode TENG, which, based on first-order lumped circuit theory, models the system as a series connection of two capacitors and a water resistor, and can be readily adapted to other liquid-solid mode TENGs [37]. Zhou et al. enhanced the performance of liquid-solid-based TENG by incorporating an end electrode into the traditional tube-based design, successfully achieving the volume effect and increasing the output voltage by approximately 40-fold to 240 V [38]. Furthermore, in 2024, Zhang et al. introduced the concept of the space volume effect in water-based TENGs, significantly enhancing output performance, with the open-circuit voltage ($V_{oc}$), short-circuit current ($I_{sc}$), and transfer charge ($Q_{sc}$) increasing by 3.5, 2.3, and 2 times, respectively [39]. Recently, Zhang et al. introduced a wave-driven closed polytetrafluoroethylene tube TENG, enhancing the conventional tank car model by utilizing the principle of interface charge transfer [40]. These advancements enabled the device to achieve an output current of 900 mA, a voltage of 150 V, and a power output of 17.74 mW. While the innovative designs of water-based solid-liquid triboelectric nanogenerators (SL-TENGs) have significantly advanced the field, several critical challenges persist, hindering their broader adoption and industrial scalability. Notably, these designs often depend on intricately fabricated plastic structures, metal electrodes, and robust sealing mechanisms, which require precise engineering and specialized manufacturing processes. Furthermore, the utilization of unique materials, such as deionized water, adds an additional layer of complexity and cost, posing significant obstacles to large-scale production and commercialization. Beyond these manufacturing and material challenges, the practical application of water-based SL-TENGs remains in its nascent stages. Although there has been progress in demonstrating the functional capabilities of these devices, their application potential is yet to be fully realized. Further research is essential to enhance their performance, reliability, and versatility, thereby unlocking their full value in practical scenarios. The development of more efficient designs and the exploration of novel materials and integration techniques will be pivotal in overcoming these challenges and driving the widespread adoption of SL-TENG technology in various fields.

Herein, we designed a highly robust and efficient water cup triboelectric nanogenerator (WC-TENG) that delivers high output performance using only drinking water and plastic cups, eliminating the need for metal electrodes and effectively addressing corrosion issues in generator components. The WC-TENG is composed of three nested water cups, with drinking water injected between each layer. The outer water layer functions as a conductive electrode, while the inner water layer simultaneously serves as both a flowing triboelectric material and a conductive electrode. Notably, the water cup's hydrophobic properties and its role as a negative triboelectric material enable it to form an effective triboelectric pair with drinking water. The three nested water cups function not only as supporting structures and triboelectric layers but also as mechanical driving components for transmitting external mechanical excitations. Experimental results demonstrate that at a working frequency of 2 Hz, the WC-TENG achieves an $V_{oc}$ of 249.71 V, a $I_{sc}$ of 4.21 μA, and a $Q_{sc}$ of 188.85 nC. The device also exhibits long-term stability and reliability, consistently maintaining a voltage output of approximately 250 V over a duration of 1500 s. The WC-TENG consistently delivers stable electrical output, demonstrating reliable performance even under high humidity conditions. Additionally, its performance improves with increased temperature, as indicated by higher voltage and charge outputs in a high-temperature environment compared to room temperature, likely due to enhanced electron transfer efficiency. Increasing the water volume in the WC-TENG enhances the triboelectric effect, resulting in higher voltage, current, and charge outputs. Conversely, the addition of ethanol initially hinders electron transfer, reducing the device's performance. However, as the ethanol volatilizes, the device gradually recovers its original charge state, demonstrating the WC-TENG's potential as a selective sensor for ethanol detection.

## 2. Experiment

### 2.1. Materials.

In this design, the WC-TENG is fabricated through an extremely simple and low-cost preparation process. The following materials were utilized in the fabrication and assembly of the WC-TENG. Three water cups fabricated by using polypropylene (PP) are utilized in the WC-TENG, serving as structural support, triboelectric layers, and mechanical transmission components within the device assembly. Drinking water was employed as the triboelectric material within the WC-TENG and conductive materials. Meanwhile, the drinking water was injected into the cups to facilitate the contact and separation process required for mechanical energy harvesting. Hot melt adhesive was utilized to securely seal the cups, preventing water leakage and ensuring the structural stability of the WC-TENG during operation. Wires were used to establish the electrical connections between the cups and the external circuit, allowing for the measurement of the generated electrical output.

### 2.2. Fabrication process of WC-TENG.

The fabrication process of WC-TENG is illustrated in Fig. 1(a1-a6). The process begins with the preparation of the individual components, water cups and drinking water. The fabrication process starts by taking a single water cup, which serves as the base structure for the WC-TENG, as shown in Fig. 1(a1). A second water cup is installed into the first cup, creating an intermediate layer, and this setup begins the formation of the triboelectric layers required for the WC-TENG device's operation, as illustrated in Fig. 1(a2). In Fig. 1(a3), drinking water is injected into the interlayer formed by two water cups to form the conductive electrode. Once the drinking water is added, the rim of the water cups is sealed using a hot melt adhesive to ensure no leakage occurs, as shown in Fig. 1(a4). Next, pour drinking water into the second cup as a liquid frictional electric material and conductive electrode, as illustrated in Fig. 1(a5). Finally, the third water cup is then installed on top of the designed structure, finalizing the assembly of the WC-TENG device, as illustrated in Fig. 1(a6). This water cup serves as the driving components in WC-TENG design. The fully assembled WC-TENG now consists of three stacked cups, each contributing to the triboelectric effect when the device is in operation. The water-filled spaces between the cups act as the active layers for energy generation. This stepwise assembly process highlights the simplicity and efficiency of the WC-TENG fabrication, emphasizing its potential for low-cost and scalable production. Wires were used to establish the electrical connections between the cups and the external circuit, allowing for the measurement of the generated electrical output. Fig. 1(b1, b2) present cross-sectional views of the fully assembled WC-TENG device, highlighting the detailed configuration of the output circuit. Fig. 1(c1) displays the materials used for the WC-TENG fabrication, including hot melt adhesive, drinking water, electrical wires, and water cups. Fig. 1(c2) shows the completed WC-TENG device with the wires connected, alongside an scanning electron microscope (SEM) image revealing the surface microstructure of the water cup.

**2.3. Characterization and measurement.**

The fully automatic mechanical exciter system, such as signal generator, power amplifier, and mechanical exciter, can provide the controllable movement required for WC-TENG device to work. The Keithley System 6517B is used to measure the $V_{oc}$ and $Q_{sc}$ of WC-TENG, and the low-noise current preamplifier (Model SR570) is used to measure the $I_{sc}$ of WC-TENG. Medical syringes are utilized to precisely administer liquid injections for the WC-TENG, while thermometers and hygrometers monitor the ambient temperature and humidity. A spray bottle is employed to elevate the humidity levels in the environment surrounding the WC-TENG, and a hair dryer is used to increase the ambient temperature.

# 3. Results and discussion

**3.1. The working mechanism of WC-TENG.**

As depicted in this Fig. 1(d), the entire apparatus is analysed from a circuitry perspective, where the resistance value of the water located in the interlayer between two water cups is denoted as $R_1$, the resistance value of the complete external circuit is denoted as $R_2$, and the resistance value of the water flow located in the the second water cup is denoted as $R_3$. Furthermore, the water cup layer is considered to act as the capacitor $C_{cup}$, and the electric double layer formed between drinking water and water cup layer is referred to as $C_t$. Considering that the thickness of the water cup significantly exceeds that of the electric double layer, the size of $C_{cup}$ in comparison to $C_t$ is negligible. The process involves the transfer of positive and negative ions in the drinking water to the surface of the water electrode and the water cup layer surface, respectively, effectively charging the capacitors $C_{cup}$ and $C_t$, leading to the electrical output. Fig. 1(d) illustrates the working mechanism of WC-TENG during its operation, which involves a cyclic process of contact electrification and charge transfer induced by mechanical movements. As illustrated in Fig. 1(d1), after continuous contact separation between the water cup layer and the drinking water, negative charges will be distributed on the surface of the water cup layer. When the flowing water is at the bottom, the drinking water electrode located in the interlayer between the first cup and the second cup will generate a positive charge layer corresponding to the negative charges distributed on the water cup layer surface. When the third cup moves downwards, the water flow will start to move upwards, causing the water flow to come into contact with the surface of the second cup, as shown in Fig. 1(d2). Cations in the water flow will shield the negative charges on the surface of the water cup layer, causing current to be generated in the circuit. Until the water flow completely covers the cup layer, no current can be generated in the external circuit, as illustrated in Fig. 1(d3). In Fig. 1(d4), when the external force acting on the third cup is released, the water flow will begin to flow towards the bottom, causing a reverse current to be generated in the external circuit.

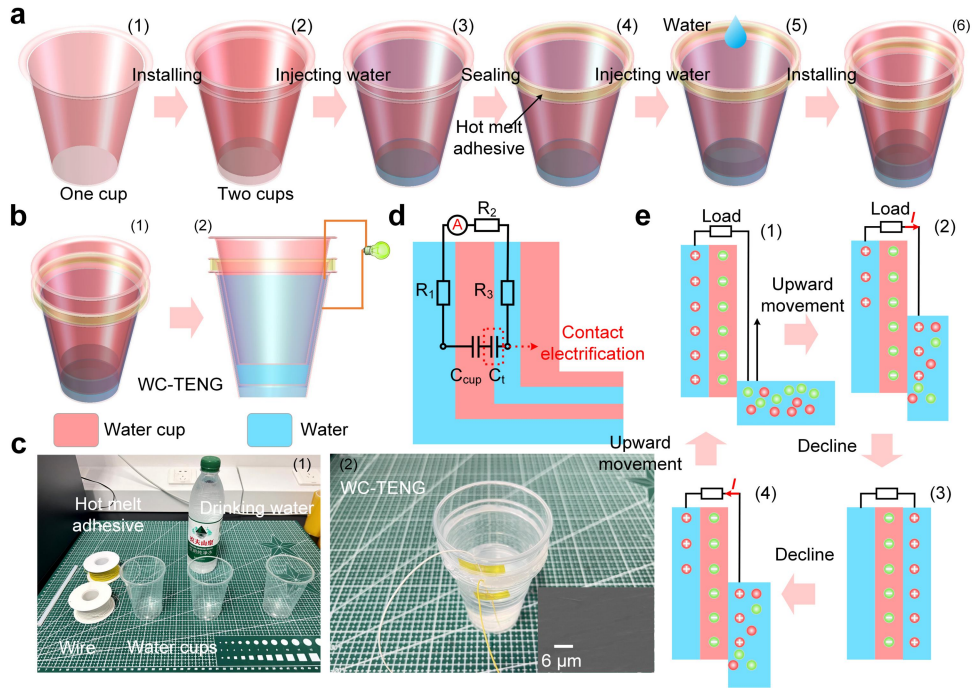

Fig. 1. Schematic illustration and assembly process of the WC-TENG. (a) Step-by-step fabrication process of WC-TENG. (b1, b2) Fully assembled WC-TENG and its connection to an external circuit for electrical output. (c1, c2) Materials used for WC-TENG fabrication, including water cups, wires, and hot melt adhesive, along with the completed device. (d) Circuit diagram of the WC-TENG during operation. (e) Working principle of the WC-TENG.

### 3.2. The electrical performance characterization of the WC-TENG.

In Fig. 2(a1), The experimental apparatus includes a mechanical vibrating system designed to simulate vertical motions, which drive the contact and separation cycles in the WC-TENG. The setup is equipped with a digital oscilloscope for real-time data acquisition, an SR570 low-noise current preamplifier for measuring $I_{sc}$, and a Keithley 6517B electrometer to measure electrical parameters such as $V_{oc}$ and $Q_{sc}$. The WC-TENG device is securely mounted on the vibrating system, with the connections to the measurement instruments clearly shown. As presented in Fig. 2(a2, a3), a close-up views of the WC-TENG's solid-liquid interface during operation, illustrating the critical stages of separation and contact between the water flow and the inner cup surface. Fig. 2(b) represents the WC-TENG connected to external measurement devices, with a voltmeter and ammeter positioned to capture the voltage and current outputs during operation. Under 2 Hz mechanical motion excitation, the $V_{oc}$ waveform generated by the WC-TENG during operation is shown, with a peak value reaching approximately 249.71 V, according to results in Fig. 2(c). The waveform demonstrates a consistent stabilize pattern, indicating stable energy harvesting performance over time. The periodic nature of the waveform corresponds directly to the cyclic contact-separation process induced by the mechanical vibrations. The $I_{sc}$ of WC-TENG is plotted over the same working frequency, displaying a peak value of approximately 4.21 µA, as illustrated in Fig. 2(d). The $I_{sc}$ waveform shows sharp peaks corresponding to the moments of maximum contact between the water and the cup surfaces, followed by gradual declines as the water flow separates from the surface. In Fig. 2(e), the accumulated $Q_{sc}$ of WC-TENG indicates a peak value of approximately 188.85 nC. The long-term stability of the voltage output of WC-TENG is demonstrated over a 1500 s period, with a consistent output of approximately 250 V. This plot indicates the durability and reliability of the WC-TENG for sustained energy harvesting, with minimal degradation in performance over extended periods of operation.

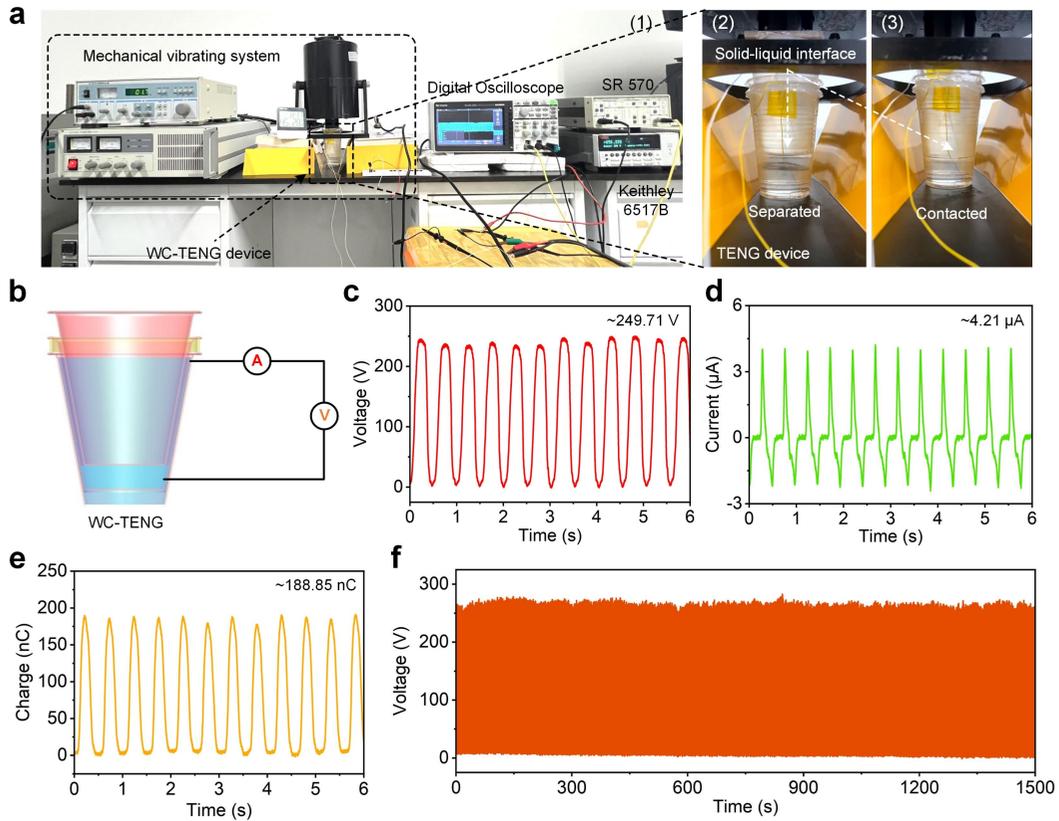

Fig. 2. Experimental setup and performance characterization of the WC-TENG. (a1) The picture of mechanical vibrating system. (a2, a3) The close-up views of the solid-liquid interface during operation. (b) Schematic of the WC-TENG circuit with voltage and current measurement. The (c) $V_{oc}$, (d) $I_{sc}$, and (e) $Q_{sc}$ of WC-TENG device under 2 Hz working frequency. (f) Long-term voltage stability of WC-TENG.

Fig. 3(a-c) illustrate the $V_{oc}$, $I_{sc}$, and $Q_{sc}$ of the WC-TENG when subjected to mechanical vibrations at different frequencies of 2 Hz, 3 Hz, 4 Hz, and 5 Hz. Fig. 3(a) illustrates the output voltage of the WC-TENG under varying mechanical frequencies, specifically at 2 Hz, 3 Hz, 4 Hz, and 5 Hz. The voltage output shows distinct sinusoidal waveforms corresponding to each frequency. Despite the increase in frequency, the Voc peak-to-peak value of WC-TENG remain consistent and stable, indicating that the WC-TENG can maintain a reliable voltage output across different operational frequencies. The $I_{sc}$ of WC-TENG follows a increasing trend, with peak values increasing from around 1.64 µA at 2 Hz to approximately 2.96 µA at 5 Hz. Fig. 3(c) presents the $Q_{sc}$ of the WC-TENG under the same set of mechanical frequencies: 2 Hz, 3 Hz, 4 Hz, and 5 Hz. Similar to the voltage output, the charge output exhibits a stable and periodic waveform across the different frequencies. Fig. 3(d-f) demonstrate the electrical output of WC-TENG at varying maximum separation distances between the solid -liquid interface, specifically 0.3 cm, 0.6 cm, 0.9 cm, 1.2 cm, and 1.5 cm. The $V_{oc}$ of WC-TENG shows a significant increase with greater separation distances, starting from about 99.39 V at 0.3 cm to nearly 297.16 V at 1.5 cm, as illustrated in Fig. 3(d). This trend suggests that a larger separation distance enhances the triboelectric effect, resulting in higher voltage generation. In Fig. 3(e), the $I_{sc}$ of WC-TENG increases correspondingly, with peak values rising from approximately 0.98 µA at 0.3 cm to about 3.03 µA at 1.5 cm. The $I_{sc}$ is directly related to the separation distance, with larger distances facilitating more substantial current generation. The accumulated $Q_{sc}$ also increases with separation

distance, starting from about 44.82 nC at 0.3 cm to over 190.24 nC at 1.5 cm, indicating that greater separation distances lead to more efficient charge generation and accumulation, as shown in Fig. 3(f). Fig.3(g-i) explore the influence of environmental humidity on the WC-TENG's output. Fig. 3(g) shows the experimental setup where the humidity is increased by spraying water, and the output performance is monitored. In Fig. 3(h), the voltage output of WC-TENG remains stable with no significant fluctuations observed, suggesting that the WC-TENG operates reliably even under varying humidity conditions. Similarly, the accumulated charge remains stable over time, with no noticeable fluctuations, indicating that the WC-TENG's performance is not adversely affected by changes in humidity, as illustrated in Fig. 3(i).

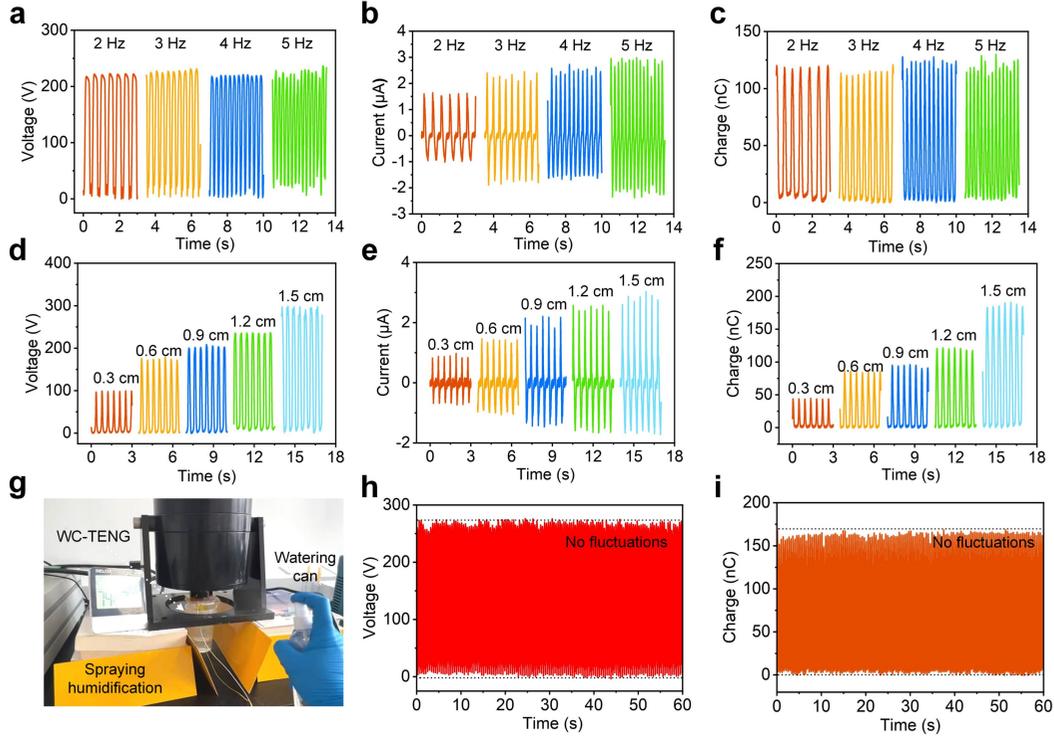

Fig. 3. Output performance of the WC-TENG under varying conditions. The (a) $V_{oc}$, (b) $I_{sc}$, and (c) $Q_{sc}$ of WC-TENG at different mechanical frequencies (2 Hz, 3 Hz, 4 Hz, and 5 Hz). The (d) $V_{oc}$, (e) $I_{sc}$, and (f) $Q_{sc}$ of WC-TENG at varying solid-liquid interface separation distances (0.3 cm to 1.5 cm). (g) The picture of measurement system for WC-TENG in high humidity environment. The (h) $V_{oc}$ and (i) $Q_{sc}$ of WC-TENG under high humidity environment.

Fig. 4(a-c) illustrates the WC-TENG's output when operating in a room temperature environment of approximately 25°C. Fig. 4(a) shows the WC-TENG positioned under the vibrating mechanism with a hygrometer indicating the environmental conditions. To assess the impact of ambient temperature on the WC-TENG's output performance, we used a hair dryer to raise the temperature of the working environment and measured the WC-TENG's output both before and after the temperature increase. From the results in Fig. 4(b), the $V_{oc}$ of the WC-TENG under these conditions is illustrated with a peak-to-peak value of approximately 159.19 V. In Fig. 4(c), the corresponding $Q_{sc}$ of WC-TENG is depicted, with a peak-to-peak value of approximately 90.64 nC. Furthermore, Fig. 4(d-f) presents the performance of the WC-TENG in a high-temperature environment of 35.1°C, achieved using a hair dryer. Fig. 4(d) shows the WC-TENG setup under the high temperature environment of 35.1°C. This comparison aims to highlight the differences in performance between high and low temperature scenarios. In Fig. 4(e), the $V_{oc}$ of WC-TENG in this higher temperature environment exhibits a peak-to-peak value of approximately 172.42 V, which is slightly higher than that observed under low temperature of 25°C, indicating that higher temperature may enhance the triboelectric effect, leading to higher voltage generation. The charge output also increases in higher temperature, with a peak-to-peak value of approximately 104.61 nC. This further supports the observation that increased temperature levels positively influence the WC-TENG's performance, likely due to enhanced electron transfer efficiency in high-temperature environments. Fig. 4(g) shows a schematic of the WC-TENG connected to a rectifier circuit for converting the alternating current (AC) output of the TENG into direct current (DC). Fig. 4(h) presents the voltage output across capacitors of varying capacitances (2 μF, 4.7 μF, and 9.4 μF) over time. The results demonstrate that the voltage across the capacitor increases more slowly with higher capacitance, indicating that the stored energy builds up progressively as the capacitance increases. The highest voltage is observed with the smallest capacitor, showing a rapid increase to a steady state, while larger capacitors take longer to charge but can store more energy overall. This research demonstrates the impact of environmental factors, specifically temperature, on the output performance of the WC-TENG. By analyzing the WC-TENG device's response to varying conditions, the study highlights the WC-TENG's potential for stable energy harvesting in diverse real-world environments, as well as its adaptability for practical applications where environmental fluctuations are common. Fig. 4(i) demonstrates the practical application of the WC-TENG by powering a series of light-emitting diodes (LEDs). The WC-TENG connected to LEDs, where the generated electrical energy is used to light up multiple LEDs, indicating successful energy harvesting and conversion. This highlights the WC-TENG's capability to directly power small electronic devices, demonstrating its potential for use in low-power applications and sustainable energy systems.

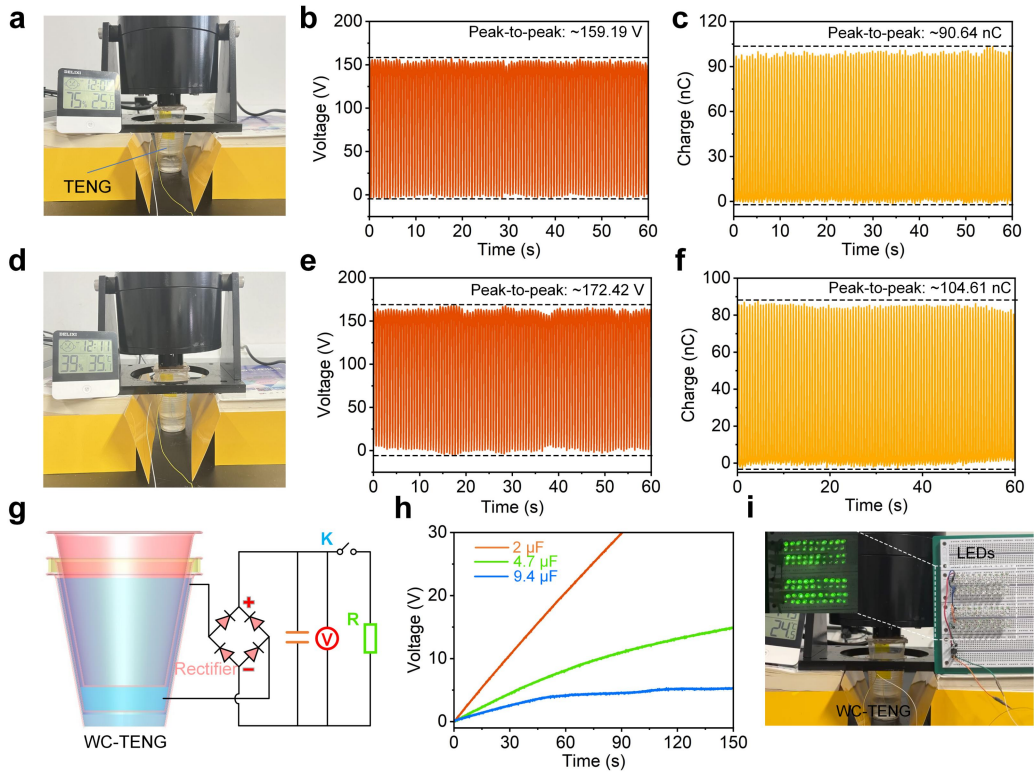

Fig. 4. Performance and application of the WC-TENG under varying environmental temperatures. (a-c) The WC-TENG output in a room temperature environment of approximately 25°C, showing stable $V_{oc}$ (~159.19 V) and $Q_{sc}$ (~90.64 nC) outputs. (d-f) The WC-TENG output in a high-temperature environment of approximately 35.1°C, with increased $V_{oc}$ (~172.42 V) and $Q_{sc}$ (~104.61 nC) outputs. (g) Schematic of the WC-TENG connected to a rectifier circuit for DC output.

(h) Voltage output across capacitors of varying capacitance (2 µF, 4.7 µF, and 9.4 µF). (i) Demonstration of the WC-TENG powering 40 commercial LEDs.

Fig. 5(a) shows the addition of drinking water to the WC-TENG, which is used to adjust the water volume inside the WC-TENG device. This process directly influences the triboelectric interaction between the water and the inner surface of the water cup. Fig. 5(b) illustrates the voltage output of the WC-TENG before and after the addition of drinking water. Initially, the voltage output exhibits a peak-to-peak value of 29.04 V. Following the addition of drinking water, the peak-to-peak voltage increases to 52.04 V, representing a change amplitude of 44%. This demonstrates that increasing the water volume enhances the triboelectric effect, thereby improving the WC-TENG device's voltage output. Fig. 5(c) illustrates the $V_{oc}$ of WC-TENG when different volumes of water (1 mL, 2 mL, 3 mL, and 4 mL) are added. The $V_{oc}$ increases progressively with the volume of water, with 1 mL generating the lowest voltage and 4 mL generating the highest, indicating a direct correlation between water volume and voltage output. Similarly, the current output is shown in Fig. 5(d) to increase with water volume. The current waveform demonstrates consistent periodicity, with higher water volumes leading to greater current generation. The highest current is observed at 4 mL, illustrating the positive effect of increased water volume on the device's performance. The $Q_{sc}$ output in Fig. 5(e) follows the same trend as voltage and current, with increasing $Q_{sc}$ accumulation as water volume increases. The consistent rise in charge with higher water volumes further confirms the enhancement of the triboelectric effect with more water. Fig. 5(f) illustrate the process of adding ethanol to the WC-TENG and the subsequent impact on the charge generation and recovery mechanism. Initially, before the addition of alcohol, drinking water interacts with the cup, causing the inner surface of the cup to carry a negative charge, as shown in Fig. 5(f1). Next, add a certain amount of alcohol to the drinking water, as shown in Fig. 5(f2). Due to the mutual solubility of alcohol and water, alcohol will rapidly diffuse and sweep into the drinking water. Meaningfully, when drinking water containing alcohol moves upwards, the alcohol comes into contact with the inner surface of the water cup, hindering the contact between cations in the water and the inner surface of the water cup, thereby reducing the output performance of WC-TENG. Multiple work cycles follow, during which the ethanol continues to interact with the WC-TENG device's surface. Over time, the ethanol begins to volatilize, resulting in the recovery of the device's original charge state, as shown in Fig. 5(f3, f4). Fig. 5(g) illustrates the voltage recovery rate after adding different volumes of ethanol (0.1 mL and 2 mL). For 0.1 mL, the initial voltage is 202.41 V, with a recovery rate of 92.56%, while for 2 mL, the initial voltage is 216.37 V, with a lower recovery rate of 76.44%. This indicates that smaller ethanol volumes result in a higher recovery rate, as less ethanol is volatilized. Fig. 5(h) shows the recovery rate as a function of ethanol volume. As the volume increases from 0.1 mL to 2 mL, the recovery rate gradually decreases, suggesting that higher ethanol volumes hinder the device's ability to return to its original charge state. This indicates the sensing function of WC-TENG in monitoring ethanol, as well as the positive correlation between WC-TENG and ethanol volume.Hence, the WC-TENG can potentially be used as a selective sensor for detecting ethanol or alcohol concentrations. The interaction between ethanol and the WC-TENG device's surface affects the triboelectric output, providing a measurable signal that correlates with ethanol presence.

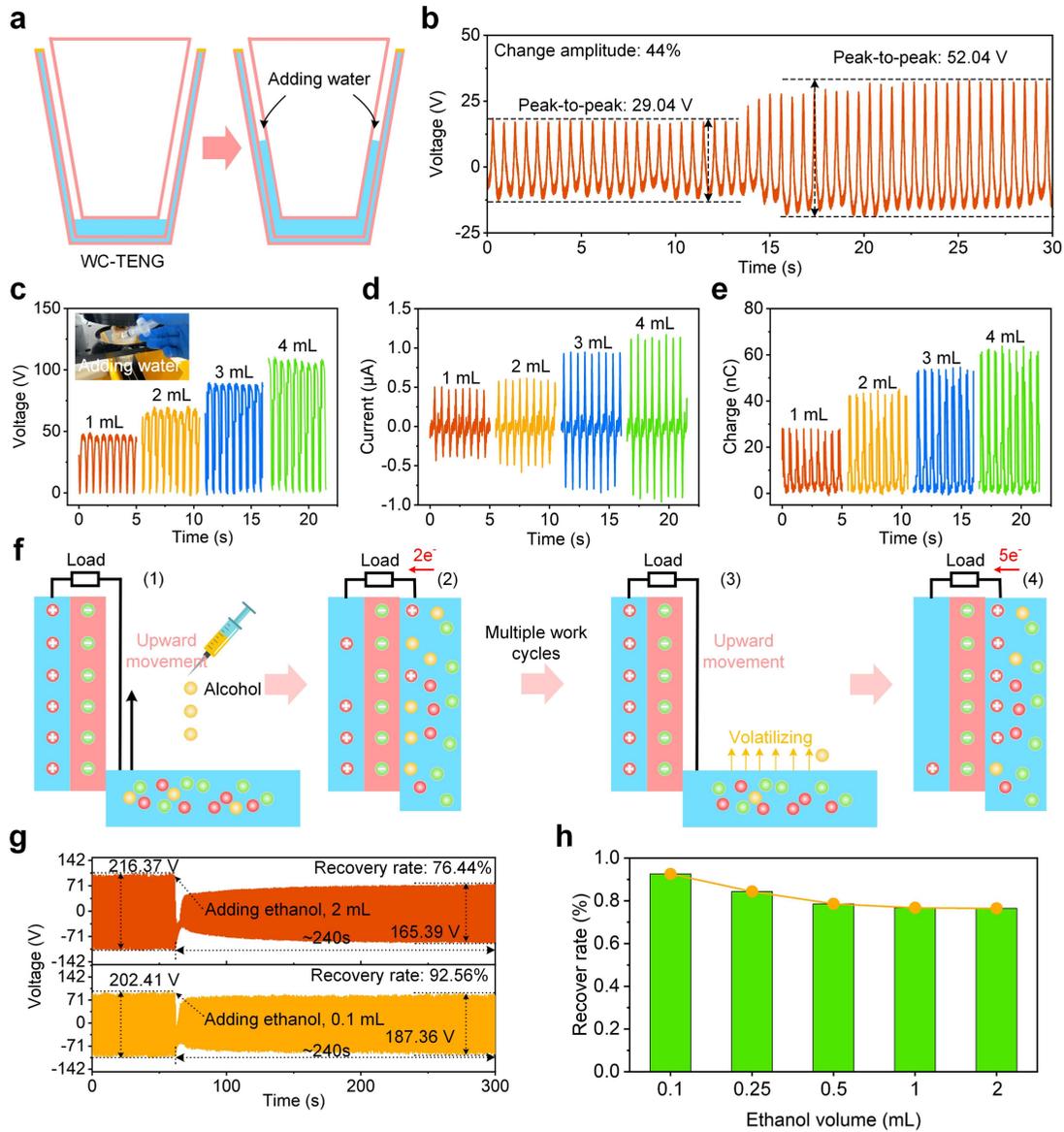

Fig. 5. Effect of water and ethanol addition on the output performance of the WC-TENG. (a) Schematic of the WC-TENG with added water. (b) Output voltage of WC-TENG before and after water addition. (c-e) Output performance with varying water volumes (1 mL, 2 mL, 3 mL, and 4 mL). (f1-f4) Schematic of the output voltage of WC-TENG and recovery mechanism upon ethanol addition. (g) Voltage recovery after adding different volumes of ethanol (0.1 mL and 2 mL). (h) Recovery rate of output voltage as a function of ethanol volume.

## 4. Conclusion

In summary, a highly robust and efficient WC-TENG that delivers high output performance using only drinking water and plastic cups, eliminating the need for metal electrodes and effectively addressing corrosion issues in generator components, was developed. Notably, the water cup's hydrophobic properties and its role as a negative triboelectric material enable it to form an effective triboelectric pair with drinking water. The three nested water cups function not only as supporting structures and triboelectric layers but also as mechanical driving components for transmitting external mechanical excitations. Experimental results indicate that, at a working frequency of 2 Hz, the WC-TENG achieves a $V_{oc}$ of 249.71 V, a $I_{sc}$ of 4.21 μA, and a $Q_{sc}$ of 188.85 nC. The device exhibits long-term stability and reliability, consistently maintaining a stable output over 1500 s. The WC-TENG demonstrates stable electrical output even under high humidity conditions and shows improved performance at elevated temperatures, likely due to enhanced electron transfer efficiency. Increasing the water volume within the WC-TENG enhances the triboelectric effect, leading to higher voltage, current, and charge outputs. In contrast, the introduction of ethanol initially hinders electron transfer, reducing the device's performance; however, as the ethanol volatilizes, the WC-TENG gradually recovers its original charge state, highlighting its potential as a selective sensor for ethanol detection.